\documentclass[aps,prd, superscriptaddress,preprintnumbers,nofootinbib]{revtex4}
\usepackage{amsmath,amsfonts,amssymb,amsthm,amstext,amscd,graphicx}
\usepackage[all]{xy}
\usepackage[colorlinks=true]{hyperref}

\def\be{\begin{equation}}
\def\ee{\end{equation}}
\def\arr{\begin{array}{rll}}
\def\ea{\end{array}}
\def\bea{\begin{eqnarray}}
\def\eea{\end{eqnarray}}

\def\N2{$N{=}2$}

\newcommand{\nnr}{\nonumber \\}
\newcommand{\nn}{\nonumber}
\def\rh{\rho_0}
\def\m{\tilde{m}}
\newcommand{\dg}{{{n}}}

\newcommand{\eps}{\varepsilon}
\newcommand{\env}[1]{{\tens{e}_{#1}}}        
\newcommand{\A}[1]{A^{\!(#1)}}                 
\newcommand{\ehf}[1]{{\hat{\tens{e}}^{#1}}}  
\newcommand{\enf}[1]{{\tens{e}^{#1}}}        
\newcommand{\ezf}{{\hat{\tens{e}}^{0}}}      
\newcommand{\ehv}[1]{{\hat{\tens{e}}_{#1}}}  
\newcommand{\ezv}{{\hat{\tens{e}}_{0}}}      

\def\l3{\ell_3}

\def\a{\alpha}

\newcommand{\tens}[1]{{\boldsymbol{#1}}}

\begin{document}
\preprint{IPM/P-2018/050}
\vskip 1 cm
\title{{\Large{Hidden symmetries of near-horizon extremal }}\\ Kerr-AdS-NUT geometries}
\author{S. Sadeghian}
\email{ssadeghian@ipm.ir }
\affiliation{{School of Physics, Institute for Research in Fundamental Sciences (IPM),}\\ P.O.Box 19395-5531, Tehran, Iran}

\begin{abstract} 
We study the hidden symmetries, the symmetries associated with Killing tensors, of the near-horizon geometry of odd-dimensional Kerr-AdS-NUT black holes in two limits: generic extremal and extremal vanishing horizon (EVH) limits. Starting from a Kerr-AdS-NUT black hole in ellipsoidal coordinates which admit integrable geodesic equations, we obtain the near-horizon extremal and EVH geometries and their principal and Killing tensors by taking the near-horizon limit. We explicitly demonstrate that geodesic equations are separable and integrable on these near-horizon geometries. We also compute the constants of motion and read the Killing tensors of these near-horizon geometries from the constants of motion. As we expected, they are the same as the Killing tensors given by taking the near-horizon limit.
\end{abstract}
\maketitle
%

\setcounter{footnote}0

\section{Introduction}
The exact symmetries in a general relativity framework are usually known as isometries that are given by Killing vectors. However, symmetries of a given metric can be generated by Killing tensors, as well. In this case, they are called \emph{hidden symmetries}, as they are not manifested in the isometries. In some special cases, hidden symmetries reduce to the isometries when Killing tensors can be trivially written as the product of Killing vectors. 

The symmetries of a metric are reflected in the motion of probe particles on that background metric such that each of the Killing vectors or tensors gives a constant of motion. If the number of (independent) constants of motion is equal to the degrees of freedom of the probe particle, its equations are integrable. If it has more independent constants of motion, the system is superintegrable.

The Killing tensors of a four-dimensional Kerr black hole and its generalization to a d-dimensional Kerr-AdS-NUT metric have been studied in Refs. \cite{Carter:1968rr,Kubiznak:2008qp}. These symmetric second-rank tensors can be written as a contraction of two (antisymmetric) Killing-Yano tensors. In some sense, a Killing-Yano tensor is the "square-root" of a Killing tensor. The Hodge dual of this Killing-Yano tensor is a closed conformal Killing two-form, called \emph{principal tensor}\cite{Frolov:2007nt}. Using the eigenvectors of the nondegenerate principal tensor, one can find a coordinate basis in which geodesics, Klein-Gordon, Dirac and Maxwell equations are separable in the probe limit\cite{Krtous:2006qy, Kubiznak:2007kh, Cariglia:2011qb,Frolov:2006pe,Lunin:2017drx}. (For a complete review, see Ref.\cite{Frolov:2017kze}.)

On the other hand, the near-horizon extremal geometry of a d-dimensional Kerr-AdS-NUT black hole is well understood. Similar to the other near-horizon extremal geometries of stationary black holes that show some universal properties including the attractor mechanism\cite{Astefanesei:2006dd,Astefanesei:2007bf} and symmetry enhancement\cite{NHEG-general,NHEG-2}, its isometry group contains the $SO(2,1)$ group. Using this, one can describe the particle dynamics with conformal mechanics\cite{conformal-mechanics-BH-1,conformal-mechanics-BH-2}. Moreover, there are special extremal black holes for which the symmetry enhances more in their near-horizon geometries. They are called extremal vanishing horizon (EVH) black holes \cite{NHEVH-1} and have been studied in Refs. \cite{NHEVH-MP, Sadeghian:2017bpr}. In these cases, we find a common $SO(2,2)$ isometry group in the near-horizon EVH geometries\cite{NHEVH-three-theorems}.

In this paper, we try to answer this question: Are the hidden symmetries enhanced in the near-horizon limit? This has been questioned in Ref. \cite{Mitsuka:2011bf} in four dimensions. Here, we start by studying the near-horizon geometry of odd-dimensional Kerr-AdS-NUT black holes in both extremal and EVH limits.\footnote{Such black holes can become EVH black holes only in odd dimensions\cite{Sadeghian:2017bpr}.} Knowing the fact that the geodesic equations on black hole geometry are separable in ellipsoidal coordinates, we take the near-horizon limit in this coordinate system. The Killing tensors and their reduction to Killing vectors of near-horizon geometries has been studied in Refs. \cite{Rasmussen:2010rw,Chernyavsky-Xu,Kolar:2017vjl} only in the extremal case. Here, we extend it to the EVH case, as well. Also, we find the principal tensor of these near-horizon extremal/EVH geometries. Following the analysis in Refs. \cite{Hakobyan:2017qee,Demirchian:2017uvo,Demirchian:2018xsk}, we study the separability of timelike geodesic equations on near-horizon extremal/EVH geometries of Kerr-AdS-NUT black holes, explicitly. Finding the constants of motion, we infer that timelike geodesics on the corresponding background metrics are integrable.

\section{A brief review of the Kerr-AdS-NUT metric}

The metric of an odd-dimensional ($d=2n+1$) Kerr-AdS-NUT black hole in ellipsoidal coordinates \cite{general kerr} is
\bea\label{Kerr-AdS-NUT}
ds^2 &=&
\sum_{\mu=1}^{n}\left( \frac{U_\mu}{X_\mu}\, dy_\mu^2
+  \frac{X_\mu}{U_\mu}\,
\Big[ \frac{W}{(1- g^2 y_\mu^2)}\,\frac{dt}{\Xi} -
\sum_{i=1}^n \frac{a_i^2 \, \gamma_i}{(a_i^2- y_\mu^2)}\,
\frac{d \phi_i}{\varepsilon_i}\Big]^2\right) \nn\\
&&
- \frac{\prod_{k=1}^n a_k^2}{\prod_{\mu=1}^{n} y_\mu^2}\,
\Big(\frac{W}{\Xi}\,dt - \sum_{i=1}^n \gamma_i\, \frac{d \phi_i}{\varepsilon_i}
\Big)^2\,,\label{oddmetric2}
\eea

where the metric functions are \footnote{The prime on the product symbol means that the factor which makes the product vanishing is removed.}
\bea\label{functions}
&&U_\mu= {{\prod}'}_{\nu=1}^n (y_\nu^2 - y_\mu^2)\,,\quad
X_\mu = \frac{(1-g^2 y_\mu^2)}{y_\mu^2}\,\prod_{k=1}^n (a_k^2 - y_\mu^2)
+ 2M_\mu\,,\qquad W= \prod_{\nu=1}^n (1-g^2 y_\nu^2)\,,\quad \nn\\
%
&&\gamma_i= \prod_{\nu=1}^n(a_i^2-y_\nu^2)\,,\label{2n1X}\quad \quad
\varepsilon_i=a_i\,\Xi_i\,{\prod_{k=1}^n}'(a_i^2-a_k^2)\,,\quad \quad \Xi_i=1-g^2\,a_i^2\,, \qquad \quad \Xi= \prod_{i=1}^n \Xi_i\,.
\eea
Considering $y_\mu=(x_\a ,ir)$ with $\a=1,\dots,n-1$, one can see that $r$ is radial direction, and $x_\a$ and ${\phi}_i$ with $i=1,\dots,n$ are related to the angular variables.

Note that $M_n$ is just equal to the mass parameter $M$, while the remaining $M_\alpha$'s are NUT parameters, denoted by $L_\alpha$.  
$a_i$ denotes the rotation parameters.

Writing the metric in terms of $r$ coordinates, one can easily see that the horizon location is given by
\bea\label{horizon}
X_n(r=r_h)= \frac{(1+g^2 r_h^2)}{r_h^2}\,\prod_{k=1}^n (a_k^2  +r_h^2)
- 2M=0\,.
\eea
The entropy and temperature of this horizon are
\bea\label{SandT}
&&S=\frac{A_{H}}{4\, G_N} = \frac{m\, {\cal A}_{d-2} }{2\, G_N\,(1+g^2 r_h^2)}\left(\prod_{i=1}^n\frac{1}{\Xi_i}\right)\,r_h\;\,,\nnr[3pt]
&&T=\frac{\kappa_{H}}{2\pi}=\frac{1}{2\pi}\big{[}\sum_{i=1}^n
\frac{r_{h}(1+g^2\, r_{h}^2)}{r_{h}^2 + a_i^2} - \frac{1}{r_{h}}\big{]}\,,
\eea
in which ${\cal A}_{n}$ is the volume of a unit n-sphere, and $G_N$ is the d-dimension Newton's constant.

We note that $\frac{\partial}{\partial t}$ and $\frac{\partial}{\partial \phi_i}$ are the Killing directions and the horizon is generated by the Killing vector,
\bea
\xi_H=\frac{\partial}{\partial t}-\sum_{i=1}^n\Omega^i\,\frac{\partial}{\partial \phi_i}\,,
\eea
where the horizon's angular velocity along each of the $\phi_i$ directions, $\Omega^i$, is given by 
\bea
\Omega^i=\frac{a_i\,(1+g^2\,r_h^2)}{r_h^2+a_i^2}\,.
\eea
Additionally, this geometry has an n-number of second-rank Killing tensors. Killing tensors are symmetric, and a rank-$r$ Killing tensor $K^{\mu_1\cdots \mu_r}$ satisfies
\bea\label{K-eq}
\nabla^{(\nu}\,K^{\mu_1\cdots\mu_r)}=0\,.
\eea
The Killing tensors of Kerr-AdS-NUT geometry, in the coordinate system in which that metric is written, are \cite{Kubiznak:2008qp}
\bea\label{KT}
K_{(k)} &=& \sum_{\mu=1}^n\Big\{
\frac{A^{(k)}_{\mu}\,X_\mu}{U_\mu}\, \Big(\frac{\partial}{\partial y_\mu}\Big)^2 +
\frac{A^{(k)}_{\mu}\, S_\mu}{y_\mu^4\, U_\mu\, X_\mu}\Big[
\frac{\partial}{\partial t} +
\sum_{k=1}^n \frac{a_k\,(1-g^2\,y_\mu^2)}{
	(a_k^2 -y_\mu^2)}\, \frac{\partial}{\partial \phi_k}\Big]^2\Big\}\nn\\
&& -\frac{c\, A^{(k)}}{\prod_{\nu=1}^n y_\nu^2}\,
\Big( \frac{\partial}{\partial t} +
\sum_{k=1}^n \frac{1}{a_k }\, \frac{\partial}{\partial\phi_k}
\Big)^2\,,\qquad k = 0, \dots, n - 1\,,\label{oddinv2}
\eea
where $c=\prod_{i=1}^n a_i^2$ and 
\bea\label{A's}
S_\mu = \prod_{k=1}^n (a_k^2-y_\mu^2)^2\,,  \quad A_{\mu}^{(k)}=\sum_{\begin{subarray}{c} \nu_1<\nu_2\dots<\nu_k \\ {\nu_1,\dots,\nu_k\ne \mu}\end{subarray} }^{n} y_{\nu_1}^2 y_{\nu_2}^2 \dots y_{\nu_k}^2,  \quad A^{(k)}=\sum_{\nu_1<\nu_2\dots<\nu_k }^{n} y_{\nu_1}^2 y_{\nu_2}^2 \dots y_{\nu_k}^2\,.
\eea
One can simply check that $K_{(0)}$ is the inverse metric, as it trivially satisfies Eq. \eqref{K-eq}.\\
After a coordinate transformation, the metric takes a simpler form (see appendix \ref{app-a} for the details):
\bea\label{simpler}
ds^2= \sum_{\mu=1}^\dg\;\biggl[\; \frac{U_\mu}{X_\mu}\,{d y_{\mu}^{2}}
  +\, \frac{X_\mu}{U_\mu}\,\Bigl(\,\sum_{j=0}^{\dg-1} \A{j}_{\mu}d\psi_j \Bigr)^{\!2}
  \;\biggr]
  -\frac{c}{\A{\dg}}\Bigl(\sum_{k=0}^\dg \A{k}d\psi_k\!\Bigr)^{\!2}\;.
\eea
The orthonormal vierbeins ${\enf\mu,\, \ehf\mu}$ (${\mu=1,\dots,\dg}$), and ${\ezf}$ are
 \begin{equation}\label{Darbouxform}
\enf\mu = {\Bigl(\frac{U_\mu}{X_\mu}\Bigr)^{\!\frac12}}\,d y_{\mu}\;,\qquad
\ehf\mu = {\Bigl(\frac{X_\mu}{U_\mu}\Bigr)^{\!\frac12}}\,
\sum_{j=0}^{\dg-1}\A{j}_{\mu}d\psi_j\;,\qquad
\ezf = {\Bigl(\frac{c}{\A{\dg}}\Bigr)^{\frac12}}\,\sum_{k=0}^{\dg}\A{k}d\psi_k\;,
\end{equation}
and their dual vectors ${\env\mu,\,\ehv\mu,\,\ezv}$ are
\begin{equation}\label{Darbouxvec}
\env\mu = {\Bigl(\frac{X_\mu}{U_\mu}\Bigr)^{\!\frac12}}\,{\partial_{y_\mu}}\;,\qquad
\ehv\mu = {\Bigl(\frac{U_\mu}{X_\mu}\Bigr)^{\!\frac12}}\,\sum_{k=0}^{\dg-1+\eps}
{\frac{(-x_{\mu}^2)^{\dg-1-k}}{U_{\mu}}}\,{\partial_{\psi_{k}}}\;,\qquad
\ezv= \bigl(c \A{\dg}\bigr)^{\!-\frac12}\,{\partial_{\psi_{\dg}}}\;.
\end{equation}
In these coordinates, the Killing tensors \eqref{KT} are also simplified to

\bea\label{killing 2n+1}
K_{(k)}=\sum_{\mu=1}^{n}A_{\mu}^{(k)}\,\Big[\env\mu \, \env\mu +\ehv\mu \,\ehv\mu \Big]-A^{(k)}\, \ezv \,\ezv \,,
\eea

Moreover, this geometry has more rich structure that admits Killing-Yano tensors. We remind the reader that a Killing-Yano tensor of rank $q$, $Y_{\mu_1\cdots \mu_q}$, is an antisymmetric tensor and solves
\bea
\nabla_{(\rho}\,Y_{\nu)\,\mu_1\cdots \mu_{q-1}}=0\,.
\eea
It is easy to show that the contraction of two Killing-Yano tensors in this way,
\bea\label{KT-KY}
K_{\mu \nu}=Y_{\mu\, \mu_1\cdots \mu_s}\,Y_{\nu\,}^{\,\mu_1\cdots \mu_s}\,,
\eea
gives a (symmetric) Killing tensor.
The Hodge dual of a $(d-2)$-rank Killing-Yano tensor is a closed conformal Killing-Yano tensor of second rank, called a principal tensor, that satisfies 
\bea
h=\star Y\,, \qquad \nabla_\rho \, h_{\mu\,\nu}=g_{\rho \mu}\,\xi_\nu-g_{\rho\,\nu}\xi_\mu\,.
\eea
Here, $\xi$ is a primary Killing vector and is defined by
\bea
\xi_\mu=\frac{-1}{(d-1)}\,\nabla^\nu\,h_{\mu\,\nu}\,.
\eea
Since the principal tensor, $h$, is closed, a potential $b$ is locally associated with it:
\bea
h=db\,.
\eea
The importance of the principal tensor is that its orthogonal (nondegenerate) eigenvectors give a coordinate basis in which geodesic equations are separable.

The principal tensor of a Kerr-AdS-NUT black hole is
\begin{equation}\label{PCCKY}
  h = \sum_{\mu=1}^{\dg} y_\mu\, d y_\mu \wedge
    \Bigl(\sum_{k=0}^{\dg-1}\A{k}_\mu d\psi_k\Bigr)=\,
         \sum_\mu y_\mu\, \enf\mu\wedge\ehf\mu\;,
\end{equation}
and its local potential ${b}$ is
\begin{equation}\label{PCCKYpot}
b = \frac12 \sum_{k=0}^{\dg-1}\A{k+1}\,d\psi_k\;.
\end{equation}

\section{near-horizon extremal geometry}\label{app-c}

The extremal limit of this black hole is given by vanishing the temperature in Eq. \eqref{SandT}. In this case, the horizon becomes degenerate and
\bea\label{extremality}
X'_n\big|_{r=r_h}=0\,.
\eea
Note that $r_h$ is the solution to Eq. \eqref{horizon}.
The near-horizon transformations are also as follows:
\bea\label{NHtr}
r=r_h+\lambda\,r_h\, \rho\,,\quad dt=\beta\,\frac{d\tau}{\lambda}\,,\quad d\phi_i=d\varphi_i+\Omega^i\,dt\,,\quad \beta=\frac{\prod_i(r_h^2+a_i^2)}{V\,r_h^3}\,.
\eea
Applying these transformations and the constraint \eqref{extremality} to the metric \eqref{Kerr-AdS-NUT} in the $\lambda \to 0$ limit, we find the near-horizon extremal geometry,
\bea\label{nearhormetric2n+1}
 &&
 ds^2=\frac{\tilde{U}_n}{V} \Big(-\rho^2 d\tau^2+\frac{d\rho^2}{\rho^2} \Big)+\frac{c}{r_h^2 \prod_{\alpha} x_\alpha^2} \Big[\frac{2\, \rho\, \tilde{U}_n}{r_h V}\, d\tau+\sum_{i=1}^{n} \tilde{\gamma}_i \,\frac{d\varphi_i}{\varepsilon_i}\Big]^2
  \nonumber\\[2pt]
&&
 \qquad +\sum_{\alpha=1}^{n-1}\frac{\tilde{U}_\alpha}{X_\alpha}\,dx_\alpha^2+\sum_{\alpha=1}^{n-1} \frac{X_\alpha}{\tilde{U}_\alpha} \Big[\frac{2\,r_h\, \rho\,\tilde{U}_n}{(r_h^2+x_\alpha^2)\,V}\, d\tau+\sum_{i=1}^{n}\frac{a_i^2\,\tilde{ \gamma}_i}{(a_i^2-x_\alpha^2)}\frac{ d\varphi_i}{\varepsilon_i} \Big]^2\,,
 \eea
where $V=-\frac{1}{2}X''_n|_{r=r_h}$.  The tilde over a function implies that it is evaluated at $r=r_h$.
\subsection{Principal and Killing tensors}\label{KT-extremal}
The Killing vectors of this geometry include the generators of rotation along $\varphi_i$,
\bea\label{rot}
 \zeta_i=\frac{\partial}{\partial \varphi_i}\,,\quad i=1\ldots n\,,
\eea
and the generators of the $sl(2,R)$, as follows:
\bea\label{SL2R-generators}
&&\xi_1={\partial_\tau}\,,\quad\nnr [3pt]
&& \xi_2=\tau\,{\partial_\tau}-\rho\,\partial_\rho\,,\quad \nnr [3pt]
&&\xi_3=\left(\tau^2+\frac{1}{\rho^2}\right)\,{\partial_\tau}-2\rho\, \tau\,{\partial_\rho}-\sum_{i=1}^n\frac{m_i}{\rho\,V}\frac{\partial}{\partial \varphi_i}\,,
\eea
where 
\bea
m_i\equiv\frac{4\,a_i\,\Xi_i\, \prod_{j=1}^{n}(r_h^2+a_j^2)}{ r_h\,(r_h^2+a_i^2)^2}\,.
\eea
These $\xi_i$'s satisfy $sl(2,R)$ algebra:
\bea
[\xi_1,\xi_2]=\xi_1\,,\qquad [\xi_1,\xi_3]=2\,\xi_2\,,\qquad [\xi_2,\xi_3]=\xi_3\,.
\eea
It is clear that $\zeta_i$'s commute with $\xi_i$'s. 
The Casimir of $sl(2,R)$ algebra is
\bea
\mathcal{I}=\frac12(\xi_1 \,\xi_3+\xi_3 \,\xi_1)-\left(\xi_2\right)^2\,.
\eea

The nontrivial Killing tensors of this geometry have been studied earlier \cite{Chernyavsky-Xu} and are given by
\bea \label{Killingtensors}
&&
\tilde{K}_{(k)}=- \frac{\tilde{A}_n^{(k)}\,V}{\tilde{U}_n\, \rho^2} \left( \frac{\partial}{\partial \tau}-\sum_{i=1}^{n}\frac{\rho\, m_i}{2\,V} \frac{\partial}{\partial {\varphi_i}} \right)^2  +\frac{\tilde{A}_n^{(k)} V \rho^2}{\tilde{U}_n} \left(\frac{\partial}{\partial \rho}\right)^2
\nonumber\\[3pt]
&&\hspace{4mm}
+\sum_{\alpha=1}^{n-1}\frac{\tilde{A}_{\alpha}^{(k)}(x_\alpha^2+r_h^2)^2}{x_\alpha^4X_\alpha \tilde{U}_\alpha}  \Bigg[\sum_{i=1}^{n} \frac{a_i\, \Xi_i \prod_{j=1}^{n}(x_\alpha^2-a^2_j)}{(x_\alpha^2-a_i^2)(r_h^2+a_i^2)} \frac{ \partial}{\partial\varphi_i} \Bigg]^2
\nonumber\\[3pt]
&&\hspace{8mm}
 +\sum_{\alpha=1}^{n-1}\frac{\tilde{A}_\alpha^{(k)} X_\alpha}{\tilde{U}_\alpha}\left(\frac{\partial}{\partial x^\alpha}\right)^2 -
\frac{\tilde{A}^{(k)}\,r_h^4}{\tilde{A}^{(n)}}\, \Big [\sum^{n}_{i=1}c_i\, \frac{ \partial}{\partial\varphi_i} \Big]^2,
\eea
where $k=0,\dots,n-1$, and the functions $\tilde{A}_\mu^{(k)}$, $\tilde{A}^{(k)}$ are related to $A_\mu^{(k)}$, $A^{(k)}$ in Eq. (\ref{A's}) by setting $x_n=\texttt{i}r_h$. The constant $c_i$'s are
\be\label{b_i}
c_i=\frac{\Xi_i\,\prod^{n}_{j=1}a_j}{a_i\,(r_h^2+a_i^2)}.
\ee

It is worth mentioning that these Killing tensors are invariant under the rotation and $sl(2,R)$ generated by Eqs. \eqref{rot} and \eqref{SL2R-generators}, respectively:
\bea\label{rot-SL2R-inv}
\mathcal{L}_{\zeta_i}\,\tilde{K}_{(k)}=0\,,\qquad \mathcal{L}_{\xi_i}\,\tilde{K}_{(k)}=0\,,\qquad \forall\, i, k
\eea

The principal potential of four-dimensional near-horizon extremal Kerr-AdS-NUT geometry has been studied in Ref. \cite{Rasmussen:2010rw}. 
However, for the d-dimensional case, it is not clear from \cite{Chernyavsky-Xu} that this near-horizon geometry has the principal tensor or not. Apparently, the answer is negative, since the principal potential $b$ seems divergent in the near-horizon limit. Here, we will show that $b$ can still be well defined in the near-horizon limit if we use this freedom: $b$ is defined up to a shift like
\bea
b\to b+C_\mu\,dx^{\mu}\,,
\eea
with constant $C_\mu$'s, that does not affect the principal tensor $h$, as $h=db$. We will show that the term which blows up in the near-horizon limit is a constant times $dt$ and can be absorbed using this freedom.

To apply the near-horizon transformations \eqref{NHtr} to $b$ in Eq. \eqref{PCCKYpot}, we should write it in terms of $dt$ and $d\phi_i$ using the coordinate transformation given in appendix \ref{app-a}. After a shift like $C\,dt$, it results in
\bea\label{b}
b=(C+b_0)\,dt\,+\,\sum_{i=1}^{n}b_i\,d\phi_i\,,
\eea
in which $b_0$ and $b_i$ are
\bea\label{b-comp}
b_0=\frac{(-1)^{n+1}}{2\,\Xi}\,\sum_{k=0}^{n-1}A^{(k+1)}\,(-g^2)^k\,,\qquad b_i=\frac{1}{2\,\varepsilon_i}\,\sum_{k=0}^{n-1}A^{(k+1)}\,(-a_i^2)^{n-k}\,,
\eea
and $\varepsilon_i$ and $\Xi$ are defined in Eq. \eqref{functions}.
In the near-horizon limit \eqref{NHtr}, $b$ takes the form
\bea\label{NH-expansion}
\tilde{b}=\left[C+\bigg({b}_0+\sum_{i=1}^n {b}_i\,\Omega^i\bigg)\bigg|_{r_h}+\bigg({b}_0+\sum_{i=1}^n {b}_i\,\Omega^i\bigg)'\bigg|_{r_h}\lambda\,r_h\,\rho+\mathcal{O}(\lambda^2)\right]\,\beta\,\frac{d\tau}{\lambda}+\sum_{i=1}^{n}\tilde{b}_i\,d\varphi_i\,.
\eea
Here, the prime denotes a derivative with respect to the $r$ coordinate. As a result of the calculations in Appendix \ref{app-b1-2}, $\left({b}_0+\sum_i {b}_i\,\Omega^i\right)\big|_{r_h}$ is constant and can be absorbed by the appropriate $C$. Therefore, it cancels the divergent term and results in
\bea
\tilde{b}=\tilde{b}_0\,\rho\,d\tau+\sum_{i=1}^{n}\tilde{b}_i\, d \varphi_i\,,
\eea
where
\bea
\tilde{b}_0\equiv \beta\,r_h\,\sum_{k=0}^{n-1}\left[\frac{\Omega^i}{\varepsilon_i}(-a_i^2)^{n-k}-\frac{(-1)^n}{\Xi}(-g^2)^k\right]\tilde{A}^{(k)}_n\,.
\eea
We note that $\tilde{b}_0$ and $\tilde{b}_i$ are functions of $x_\alpha$ through $\tilde{A}^{(k)}_n$ and $\tilde{A}^{(k+1)}$, respectively, and do not depend on $\rho$.

Such behavior has been investigated, with an explicit example in Appendix \ref{app-b1-1}.

\subsection{Integrability of geodesic equations}

The simplest constant of motion for timelike geodesics is,
\bea
g^{ab}p_a\,p_b=-m_0^2\,.
\eea
Using the inverse metric of near-horizon geometry given by the $k=0$ of Eq. \eqref{Killingtensors}
and the projection of the Casimir element $\mathcal{I}$ onto the momentum space, we have
\bea\label{mass-1}
\frac{V}{\tilde{U}_n}\,\left(\mathcal{I}+\Big[\sum_i \frac{m_i\,p_i}{{2\,V}}\Big]^2\right)+
\frac{r_h^4}{\tilde{A}^{(n)}}\, \Big [\sum^{n}_{i=1}c_i\, p_i \Big]^2-\sum_{\a=1}^{n-1}\left[\frac{\tilde{X}_\a}{\tilde{U}_\a}\,p_{\a}^2+\sum_{i,j=1}^n \frac{M_\a^{ij}}{\tilde{U}_\a}\,p_i\,p_j\right]=m_0^2\,,
\eea
where $M_\a^{ij}$ is defined by
\bea
M_\a^{ij}\equiv\frac{(x_\alpha^2+r_h^2)^2\,\prod_{l=2}^{n}(a_l^2-x_\a^2)^2\,a_i\,\Xi_i\,a_j\,\Xi_j}{x_\a^4\,X_\a\,(a_i^2-x_\a^2)(a_j^2-x_\a^2)(r_h^2+a_i^2)(r_h^2+a_j^2)} \,.
\eea
The angular Hamiltonian, $\mathcal{E}$, which is defined by
\bea\label{agular-H}
\mathcal{E}\equiv V\,\left(\mathcal{I}+\frac{1}{4\,V^2}\Big[\sum_i m_i\,p_i\Big]^2\right)=\frac{V}{\rho^2}\left(p_0-\sum_i \frac{\rho\,m_i}{2\,V}\,p_i\right)^2-V\,\rho^2\,p_{\rho}^2\,,
\eea
can be rewritten in terms of angular variables using equation Eq. \eqref{mass-1}.

Similar to the analysis of Refs. \cite{Hakobyan:2017qee,Demirchian:2018xsk}, we see that that the Hamilton-Jacobi equations are also separable on near-horizon extremal geometry of an odd-dimensional Kerr-AdS-NUT black hole.
Using the identity Eq. \eqref{27},
the equation \eqref{mass-1} can be conveniently rewritten as

\be\label{H-J-E}	
\sum_{\mu=1}^{n}\frac{1}{{\prod_{\nu=1}^{n}}'\,(x_\nu^2-x_\mu^2)}\left[R_\mu +W_\mu\right]=0,
\ee
where
\bea\label{R&W}
&&R_n\equiv -\mathcal{E}\,,\qquad \nnr
&&R_\a \equiv X_\a\,p_\a^2+\sum_{i,j=1}^n M_\a^{ij}\,p_i\,p_j\,, \nnr 
&&W_\mu\equiv m_0^2\left(-x_\mu^2\right)^{n-1}- \frac{r_h^4}{x_\mu^2}\, \Big [\sum^{n}_{i=1}c_i\, p_i \Big]^2\,.
\eea

Recalling the identity \eqref{n-2},
we can rewrite the expression \eqref{H-J-E} in the form
\be\label{HJ-1}
\sum_{\mu=1}^{n}\frac{1}{{\prod_{\nu=1}^{n}}'\,(x_\nu^2-x_\mu^2)}\Big[R_\mu+W_\mu
	-\sum_{k=1}^{n-1}\nu_{k}\,\left(-x_\mu^2\right)^{n-1-k}\Big]=0\,.
\ee
Here, $\nu_k$'s are some arbitrary and independent constants which can be considered as constants of motion. To find the $\nu_k$'s, we should reverse the equation:
\bea\label{eq-1}
R_\mu+W_\mu=\sum_{k=1}^{n-1}\nu_{k}\,\left(-x_\mu^2\right)^{n-1-k}\,.
\eea

By multiplying it with $\frac{\tilde{A}_\mu^k}{{\prod_{\nu}}'\,(x_\nu^2-x_\mu^2)}$, summing over $\mu$, and using the identities \eqref{28}, and \eqref{29},
%
we have
\bea
\nu_k=-\frac{\tilde{A}^{(k)}_n}{\tilde{U}_{n}}\,\mathcal{E}+\sum_{\a=1}^{n-1}\frac{\tilde{A}_\a^{(k)}\,R_\a}{\tilde{U}_{\a}}-
\frac{r_h^4\,A^{(k)}}{\tilde{A}^{(n)}}\, \Big [\sum^{n}_{i=1}c_i\, p_i \Big]^2\,,\quad k=1,\ldots , n-1\,.
\eea

In addition to these $(n-1)$ constants of motion, $m_0^2$ is also a constant of motion. This can be considered as $\nu_0$ by the shift
\bea
\nu_k\to \nu_k-m_0^2\, \delta_{k,0}\,.
\eea
We note to the reader that the range of $k$ was $[1,n-1]$ initially and did not include $k=0$. However, we extended it to include $k=0$.

Recalling Eq. \eqref{R&W} for the definition of $R_\a$ and $\mathcal{E}$ from \eqref{agular-H}, we have
\bea
&&\nu_k=-\frac{V\,\tilde{A}^{(k)}_n}{{\tilde{U}_n}\,\rho^2}\left(p_0-\sum_i \frac{\rho\,m_i}{2\,V}\,p_i\right)^2+\frac{V}{\tilde{U}_n}\rho^2\,p_{\rho}^2\, \tilde{A}^{(k)}_n\,, \nnr [3pt]
&&\hspace{1cm}+\,\sum_{\a=1}^{n-1}\frac{\tilde{A}_\a^{(k)}}{\tilde{U}_\a}\Big[X_\a\,p_\a^2+\sum_{i,j=1}^n M_\a^{ij}\,p_i\,p_j\Big]-\frac{r_h^4\,A^{(k)}}{\tilde{A}^{(n)}}\, \Big [\sum^{n}_{i=1}c_i\, p_i \Big]^2\,,\
\eea
where $k$ runs over $[0, n-1]$ now.
Considering these constants, $\nu_k$, as the contraction of Killing tensors, $K_{(k)}^{\mu \nu}$, with momentum $p_\nu$,
\bea
\nu_k=K_{(k)}^{\mu \nu}\,p_\mu\,p_\nu\,,
\eea
and one can readily see that the resultant Killing tensors are the same as the Killing tensors in Eq. \eqref{Killingtensors} that we obtained by taking the near-horizon limit. For instance, the Killing tensor related to $\nu_0$ is the metric itself (since $A^{(0)}=A^{(0)}_\mu=1$).

In addition to these $n$ constants of motion made of Killing tensors, there are $n$ constants of motion associated with Killing vectors $\zeta_i$, of the form $\zeta_i^\mu\,p_\mu$, and two others from the Cartan and Casimir elements of $sl(2,R)$. As a result of Eq. \eqref{rot-SL2R-inv}, these $2n+2$ constants are Poisson commuting. However, all these constants of motion are \emph{not} independent, and there is a constraint between them:

\bea \label{reducible-K}
\sum_{k=0}^{n-1}\nu_{k} \left({r_h}^2\right)^{n-1-k}\,&=&-\,\mathcal{E}\,+ r_h^2\Big [\sum^{n}_{i=1}c_i\, p_i \Big]^2\,,
\eea
as a combination of the corresponding Killing tensors can be written in terms of Killing vectors\cite{Rasmussen:2010rw,Chernyavsky-Xu}. Altogether, geodesic equations on the near-horizon extremal geometry of a ($d=2n+1$)-dimensional Kerr-AdS-NUT black hole have a $2n+1$ \emph{independent, commuting} constant of motion; therefore, they are integrable.

\section{near-horizon EVH geometry}
In the previous section, it is assumed that the horizon area is nonzero. On the other hand, one can take a limit in which both horizon area and temperature vanish with the same rate. That is called the EVH limit. As one can readily see from the form of entropy in Eq. \eqref{SandT}, the horizon area vanishes once $r_h$ goes to zero. The EVH limit of a Kerr-AdS-NUT black hole in odd dimensions has been studied in Ref. \cite{Sadeghian:2017bpr} with more details. It is given by a limit
\bea \label{EVHscaling}
r_{h}=\rh\, \epsilon,\, \ \ \ a_1=\tilde{a}_1\, \epsilon^2, \, \, \ M=\frac{1}{2}\,\prod_{i=2}^{n}a_{i}^2+\tilde{M}\, \epsilon^2, \ \ \ \ \ \, \, \, \epsilon\rightarrow 0,
\eea
where the parameter $\m$ is given by
\bea \label{mtilde}
\tilde{M}=\frac{\rh^2}{2} \left(\frac{\tilde{a}_1^2}{\rh^4}-\lambda_3\right)\; \prod_{i=2}^{n}a_i^2\,,\qquad \lambda_3\equiv -g^2-\sum_{i=2}^n\frac{1}{a_i^2}\, .
\eea
(using the notations of Ref. \cite{Sadeghian:2017bpr} for $\lambda_3$).

To obtain the near-horizon limit of an EVH black hole, we apply the EVH limit \eqref{EVHscaling} and the following transformations to the metric in Eq. \ref{Kerr-AdS-NUT}:
\bea \label{nearhorizon}
 t=\beta_3\, \frac{\tau}{\gamma},\,\quad  r=r_{h}+\gamma\, \rho\,, \quad
\phi_1=p\;\frac{\varphi}{\gamma}\, , \quad \phi_{i}=\varphi_{i}+\Omega^i\, t\,, \quad  2\leq i \leq n \,,
\eea
where
\bea
&&\beta_3=\frac{1}{V_3}\,\prod_{k=2}^{n}a_k^2\,,\qquad p=\beta_3\, \sqrt{V_3}\,,\nnr
&& V_3=-\frac{1}{2}X_n''\Big|_{r=r_h}=(-\lambda_3)\,\prod_{i=2}^{n}a_i^2\,,
\eea
and we assume that $\epsilon \ll \gamma$ in the $\epsilon,\gamma\rightarrow 0$ limit. In this case, the near-horizon of an EVH black hole (NHEVH) reads as
\begin{gather} \label{NH_EVH}
ds^2_{NH}=\frac{\tilde{U}}{V_3}\left( -\,\rho^2d\tau^2+\,\frac{d\rho^2}{\rho^2}+\rho^2 \;d\varphi^2\right)+\sum_{\a=1}^{n-1}\left[\frac{\tilde{U}_\a}{\tilde{X}_\a}\,dx_\a^2+\frac{\tilde{X}_\a}{\tilde{U}_\a}\,\left(\sum_{i=2}^{n}\frac{a_i^2\, \tilde{\gamma}_i}{ (a_i^2-x_\a^2)\,\tilde{\varepsilon}_i}d\varphi_i\right)^2\right]\,,
\end{gather}
where $\tilde{\gamma}_i$ and $\tilde{\varepsilon}_i$ are
\bea\label{metricH}
\tilde{\gamma}_i=a_i^2\prod_{\a=1}^{n-1}(a_i^2-x_\a^2) \,,\qquad \tilde{\varepsilon}_i=a_i\,{\prod_{k=2}^{n}}'(a_i^2-a_k^2) \,.
\eea
We note that the tilde on each quantity means that it is computed in the EVH and near-horizon limit. Therefore, the metric functions are
\bea\label{def-tilde}
&&\tilde{U}\equiv\tilde{U}_n=\prod_{\a=1}^{n-1}x_\a^2\,,\qquad \quad \,  \, \tilde{X}_\a=-(1-g^2x_\a^2)\,\prod_{k=2}^{n}(a_k^2-x_\a^2)+2L_\a\,,\nnr
&& \tilde{U}_\a=-x_\a^2\,{\prod_{\beta=1}^{n-1}\,}'(x_\beta^2-x_\a^2)\,,\qquad \qquad \quad \qquad S_\a=x_\a^4\,\prod_{k=2}^{n}(a_k^2-x_\a^2)^2\,\,.
\eea
One can check that this geometry is also a solution to the pure Einstein theory.

\subsection{Principal and Killing tensors}\label{KT-EVH}

The near-horizon geometry of Kerr-AdS-NUT black hole is given in the previous section.
As is clear from Eq. \eqref{NH_EVH}, the metric includes an $AdS_3$ factor and is invariant under the $SO(2,2)$ group.
It can be viewed as two copies of $SL(2,R)$. In the coordinates
\bea
v=\tau+\varphi\,,\qquad u=\tau-\varphi\,,
\eea
the generators of these two $SL(2,R)$ are as follows:
\bea
&&H^+={\partial_v}\,,\qquad D^+=v\,{\partial_v}-\rho\,\partial_\rho\,\qquad K^+=v^2\,{\partial_v}+\frac{1}{\rho^2}\,{\partial_u}-2\rho\, v\,{\partial_\rho}\,,\nnr
&&H^-={\partial_u}\,,\qquad D^-=u\,{\partial_u}-\rho\,\partial_\rho\,\qquad K^-=u^2\,{\partial_u}+\frac{1}{\rho^2}\,{\partial_v}-2\rho\, u\,{\partial_\rho}\,,
\eea
and each of these sets satisfies $sl(2,R)$ algebra:
\bea
[H^a,D^a]=H^a\,\qquad [H^a,K^a]=2D^a\,,\qquad [D^a,K^a]=K^a\,, \quad a=+,-.
\eea
The Casimir of each copy is
\bea
\mathcal{I^\pm}=\frac12\,(H^\pm \,K^\pm+K^\pm \,H^\pm)-\left(D^\pm\right)^2\,.
\eea
One can simply check that the Casimirs are equal.

Applying the EVH limit \eqref{EVHscaling} and near-horizon limit \eqref{nearhorizon} to the second-rank Killing tensors in Eq. \eqref{KT} gives
\bea\label{NHEVHKT}
&&K_{(k)}=\tilde{A}^{(k)}\,\frac{V_3}{\tilde{U}}\left(-\frac{1}{\rho^2}\left({\partial_\tau}\right)^2+\,\rho^2\left({\partial_\rho}\right)^2+\frac{1}{\rho^2}\left({\partial_\varphi}\right)^2\right)\,,\nonumber\\
&&\qquad \, +\,\sum_{\a=1}^{n-1} \tilde{A}^{(k)}_{\a}\left(\frac{\tilde{X}_\a}{\tilde{U}_\a}\Big(\frac{\partial}{\partial x^\a}\Big)^2+\frac{S_\a}{\tilde{X}_\a\, \tilde{U}_\a}\Big(\sum_{i=2}^{n}\frac{\Xi_i}{{a_i}\,(a_i^2-x^2_\a)}\frac{\partial}{\partial \varphi^i}\Big)^2\right)\,.
\eea
One can simply check that the $k=0$ case of them is just the inverse metric of the NHEVH of a Kerr-AdS-NUT black hole in Eq. \eqref{NH_EVH}.

The principal tensor $h$ can be read from its potential $b$, defined in Eq. \eqref{PCCKYpot}, by applying the near-horizon and EVH limits from Eqs. \eqref{nearhorizon} and \eqref{EVHscaling}, and taking the transformation \eqref{transf} into account. This gives
\bea
\tilde{b}=\sum_{i=2}^{n}\tilde{b}_i\, d \varphi_i\;,
\eea
in which $\tilde{b}_i$'s are the $b_i$'s in Eq. \eqref{b-comp} that should be computed in the near-horizon EVH limit.

\subsection{Integrability of geodesic equations}

Again, we start from the simplest constant of motion for geodesics, i.e.,
\bea
g^{ab}p_a\,p_b=-m_0^2\,.
\eea
Using the inverse metric of near-horizon geometry given by the $k=0$ of Eq. \eqref{NHEVHKT}
and the projection of the Casimir element $\mathcal{I}$ onto the momentum space, we have
\bea\label{mass-simplified1}
-\frac{V_3}{\tilde{U}}\,\mathcal{I}+\sum_{\a=1}^{n-1}\frac{\tilde{X}_\a}{\tilde{U}_\a}\,p_{\a}^2+\sum_{\a=1}^{n-1} \sum_{i,j=2}^n\frac{M_\a^{ij}}{{\prod_{\beta=1}^{n-1}\,}'(x_\beta^2-x_\a^2)}\,p_i\,p_j=-m_0^2\,,
\eea
where $M_\a^{ij}$ is defined by
\bea
M_\a^{ij}\equiv\frac{(-x_\a^2)\,\prod_{l=2}^{n}(a_l^2-x_\a^2)^2}{X_\a\,(a_i^2-x_\a^2)(a_j^2-x_\a^2)}\,\frac{\Xi_i\,\Xi_j}{a_i\,a_j}\,.
\eea
The angular Hamiltonian, $\mathcal{E}$, which is defined by
\bea\label{agular-H-def}
\mathcal{E}\equiv V_3\,\mathcal{I}=V_3\,\left(\frac{1}{\rho^2}\left(p_0^2-p_{\varphi}^2\right)-\rho^2\,p_{\rho}^2\right)\,,
\eea
can be simplified using Eq. \eqref{mass-simplified1}:
\bea\label{Casimir-simplified}
\mathcal{E}=\left(\prod_{\a=1}^{n-1}x_\a^2\right)\,\left(\sum_{\a=1}^{n-1}\frac{\tilde{X}_\a}{\tilde{U}_\a}\,p_{\a}^2\,+\sum_{\a=1}^{n-1}\sum_{i,j=2}^n\frac{M_\a^{ij}}{{\prod_{\beta=1}^{n-1}\,}'(x_\beta^2-x_\a^2)}\,p_i\,p_j+m_0^2\right)\,.
\eea

Following the analysis of the separability of Hamilton-Jacobi equations in Refs. \cite{Hakobyan:2017qee,Demirchian:2018xsk} reveals that the Hamilton-Jacobi equations are also separable on the near-horizon EVH geometry of a Kerr-AdS-NUT black hole in odd dimensions.
Using the identity \eqref{27},
the angular Hamiltonian ${\cal E}$, given in Eq. \eqref{Casimir-simplified}, can be conveniently represented through

\be\label{HJO}	
\sum_{\a=1}^{n -1}\frac{1}{{\prod_{\beta=1}^{n-1}\,}'(x_\beta^2-x_\a^2)}\left(R_\a(p,x)- \frac{\mathcal{E}}{x_\a^2}\right)=0,
\ee
where
\be
R_\a(p,x)\equiv -\frac{\tilde{X}_\a}{x_\a^2}\,p_\a^2+\sum_{i,j=2}^n M_\a^{ij}\,p_i\,p_j+m_0^2\left(-x_\a^2\right)^{n-2}.
\label{24}\ee
Recalling the identity \eqref{n-2},
we can rewrite the expression \eqref{HJO} in a more useful form:
\be\label{HJ1}
\sum_{\a=1}^{n-1}\frac{1}{{\prod_{\beta=1}^{n-1}\,}'(x_\beta^2-x_\a^2)}\left(R_\a(p,x)
	-\sum_{k=1}^{n-1}\nu_{k}\,(-x_\a^2)^{n-2-k}\right)=0,\qquad \nu_{n-1}=-\mathcal{E}\,.
\ee
Here, $\nu_k$'s are some arbitrary and independent constants which can be considered as constants of motion. To find $\nu_k$'s, we should reverse the equation:
\bea\label{eq-1}
R_\a(p,x)=\sum_{k=1}^{n-1}\nu_{k}\,\left(-x_\a^2\right)^{n-2-k}\,.
\eea
By multiplication it with $\frac{\tilde{A}_\a^k}{{\prod_{\beta}\,}'(x_\beta^2-x_\a^2)}$, summation over $\a$, and using the identities \eqref{28} and \eqref{29},
%
we have
\bea
\nu_k=-\frac{\tilde{A}^{(k)}}{\tilde{A}^{(n-1)}}\,\mathcal{E}+\sum_{\a=1}^{n-1}\frac{\tilde{A}_\a^{(k)}\,R_\a}{{\prod_{\beta=1}^{n-1}\,}'(x_\beta^2-x_\a^2)}\,,\qquad k=1,\ldots , n-2\,.
\eea

This result can be rewritten 
 by substituting $\mathcal{E}$ from 
Eq. \eqref{agular-H-def}
 and noting that $\tilde{A}^{(n-1)}$ is just $\prod_{\a}x_\a^2$, as
\bea
\nu_k=\frac{V_3}{\tilde{U}}\,\left(\rho^2\,p_{\rho}^2-\frac{1}{\rho^2}\left(p_0^2-p_{\varphi}^2\right)\right)\tilde{A}^{(k)}+\,\sum_{\a=1}^{n-1}\frac{\tilde{A}_\a^{(k)}\,R_\a}{{\prod_{\beta=1}^{n-1}\,}'(x_\beta^2-x_\a^2)}\,,
\eea

In addition to these $(n-1)$ constants of motion, $m_0^2$ is also a constant of motion. This can be considered as $\nu_0$ by the shift
\bea
\nu_k\to \nu_k-m_0^2\, \delta_{k,0}\,. 
\eea
We note to the reader that the range of $k$ was $[1,n-1]$ initially and did not include $k=0$. However, we extended it to include $k=0$. Recalling Eq. \eqref{24} for the definition of $R_\a$, we have
\bea
&&\hspace{-.6cm}\nu_k=\frac{V_3}{\tilde{U}}\,\left(\rho^2\,p_{\rho}^2-\frac{1}{\rho^2}\left(p_0^2-p_{\varphi}^2\right)\right)\tilde{A}^{(k)}\,+\,\sum_{\a=1}^{n-1} \tilde{A}^{(k)}_{\a}\left(\frac{\tilde{X}_\a}{\tilde{U}_\a}\,p_\a^2+\frac{S_\a}{\tilde{X}_\a\, \tilde{U}_\a}\Big(\sum_{i=2}^{n}\frac{\Xi_i\, p_i}{{a_i}\,(a_i^2-x^2_\a)}\Big)^2\right)\,,\nnr
\eea
where $k$ now runs over $[0, n-1]$.
Considering these constants, $\nu_k$, as the contraction of Killing tensors, $K_{(k)}^{\mu \nu}$, with momentum $p_\mu$,
\bea
\nu_k=K_{(k)}^{\mu \nu}\,p_\mu\,p_\nu\,,
\eea
one can readily see that the resultant Killing tensors are the same as the Killing tensors in Eq. \eqref{NHEVHKT} that we obtained by taking the near-horizon limit. 

Similar to the constraint \eqref{reducible-K} in the extremal case, we have 
\bea
\nu_{n-1}=-\mathcal{E}\,,
\eea
and all $\nu_k$'s, are not independent of the Casimir. Therefore, we have $(2n+1)$ independent constants of motion in this case. So the geodesic equations on the near-horizon EVH Kerr-AdS-NUT geometry are also integrable and separable.



\section{Discussion and conclusion}
In this work, we studied the principal and Killing tensors of near-horizon extremal and EVH geometries of a Kerr-AdS-NUT black hole in odd dimensions. The even-dimensional case can be analyzed in a similar manner for the extremal case. Although the Killing tensors were given for the extremal case earlier \cite{Chernyavsky-Xu,Kolar:2017vjl}, we improve the discussion of hidden symmetries by introducing the principal tensor for near-horizon extremal and EVH geometries. The principal tensor is a closed form and locally accompanied by a potential.  Then, this potential is defined up to an exact form. In the near-horizon limit, we used this freedom to make the principal tensor finite. The existence of this tensor for a given metric makes geodesics, Klein-Gordon, Dirac and Maxwell field equations separable on that background metric. We explicitly showed the separability of timelike geodesic equations on the mentioned near-horizon geometries.

Finding the constants of motion associated with the geodesics, one can read the Killing tensors of the background metric. We observed that the obtained Killing tensors in this way are the same as the Killing tensors given by taking the near-horizon limit.

One may also study the Penrose process and superradiance in these spacetimes and see if there are some distinct features due to taking special limits like in Ref. \cite{Mukherjee:2018dmm}.

It is well known that the isometries enhance in the near-horizon extremal limit, and Killing vectors have $sl(2,R)$ algebra. However, there is no extra structure among the given second-rank Killing tensors and Killing vectors. Particularly, hidden symmetries (associated with the given second-rank Killing tensors) \emph{do not enhance} in the near-horizon extremal or EVH limit. This statement should be revised for the equal angular momenta or for null geodesics. 

In spite of the fact that the Casimir of $sl(2,R)$ gives an extra constant of motion for the geodesic equations on near-horizon extremal geometry, this problem is still integrable (not superintegrable) since there is a relation between the constants associated with the Killing tensors, Casimir and Killing vectors. One may expect that for the EVH case where we have $so(2,2)$ as a subgroup of the isometries, we have more constants of motion. But this is not the case, because two constants that the Casimirs of $so(2,2)$ provide are equal. Therefore, a geodesic problem on near-horizon EVH geometry of a Kerr-AdS-NUT black hole is also integrable.

These geometries have dual CFT descriptions from the AdS/CFT point of view. It would be interesting to find the meaning of the Killing tensors, hidden symmetries, and integrability of geodesics on the CFT side.
\section*{Acknowledgments}
The author is grateful to Hovhannes Demirchian, Armen Nersessian, and especially M.M. Sheikh-Jabbari for discussions during our previous collaborations. I also thank the conference on ``Gravity - New perspectives from strings and higher dimensions,'' where the project was initiated. I learned hidden symmetries from David Kubiznak there and thank him. This work is partially supported by ICTP Program Network Scheme No. NT-04.

\appendix
\section{A useful coordinate transformation}\label{app-a}
Using the transformations
\bea
\tilde{t}=\frac{t}{\Xi}\,,\qquad \tilde{\phi}_i=\frac{\phi_i}{\varepsilon_i}\,,
\eea
on the Kerr-NUT-AdS metric \eqref{Kerr-AdS-NUT} in odd dimensions ($D=2n+1$) and the definitions
\bea a_0 &=& \frac1{g}\,,\ \ \qquad \Gamma_I= \prod_{\nu=1}^n (a_I^2 -
y_\nu^2)\,,
\quad 0\le I\le n\,,\nn\\
\tilde\phi_0 &=& - g^{2n}\, \tilde t\,,\qquad X_\mu=
\frac{g^2}{y_\mu^2}\, \prod_{I=0}^{n} (a_I^2 -y_\mu^2) + 2M_\mu\,.
\eea
the metric can be written as
\be \label{intermediate-metric}
ds^2 = \sum_{\mu=1}^{n} \Big\{ \frac{U_\mu}{X_\mu}\, dy_\mu^2 +
\frac{X_\mu}{U_\mu}\, \Big( \sum_{I=0}^{n}
\frac{a_I^2 \, \Gamma_I\, d\tilde\phi_I}{a_I^2-y_\mu^2} \Big)^2\Big\} 
- \frac{(\prod_{k=1}^n a_k^2)}{(\prod_{\mu=1}^{n} y_\mu^2)}\,
\Big(\sum_{I=0}^{n} \Gamma_I\, d\tilde\phi_I
    \Big)^2\,.
\ee
By using the relations
\bea
&&\Gamma_I=(-1)^n\sum_{k=0}^n A^k (-a_I^2)^{n-k}\,,\nnr
&&(a_I^2-y_\mu)^{-1}\,\Gamma_I=(-1)^{n-1}\sum_{k=0}^{n-1} A^k_\mu (-a_I^2)^{n-k-1}\,,
\eea 
the metric \eqref{intermediate-metric} takes the simpler form \eqref{simpler} if we define
\bea\label{transf}
d\psi_k=\sum_{I=0}^n \left(-a_I^2\right)^{n-k}d\tilde{\phi}_I\,.
\eea


\section{Principal tensor of near-horizon extremal geometry} \label{app-b1}
\subsection{Case study : 5D NHEMP}\label{app-b1-1}
In this part, we restrict our attention to the $g=0$, $L_\alpha=0$, and $d=5$ case i.e., to the near-horizon extremal geometry of a five dimensional Myers-Perry black hole \cite{mp}.
This solution is described by two rotation parameters, $a_1,a_2$. After solving Eq. \eqref{horizon} for the horizon location in the extremal limit [Eq. \eqref{extremality}], we find that
\bea
r_h^2=a_1\,a_2\,, \qquad 2M=(a_1+a_2)^2\,.
\eea
The near-horizon metric is
\bea
ds^2&&\hspace{-3mm}=-{\frac {\tilde{U}{\rho}^{2}{{\it d\tau}}^{2}}{V}}+{\frac {\tilde{U}{{\it d\rho}}^{2}}{V{\rho}^{2}
}}+{\frac {{\it \tilde{U}_1} (x)}{{\it X_1}(x) }} {{\it dx}}^{2} +\frac{{a_1}^{2}\,{a_2}^{2}}{r_h^2\,x^2} \left( 2\,{\frac {\tilde{U}\,\rho\,{\it d\tau}}{V{\it r_h}}
}+{\frac {{\it \tilde{\gamma}_1} (x)\, {\it d\varphi_1}}{\varepsilon _1}}+{\frac {{\it \tilde{\gamma}_2} (x)\, {\it d\varphi_2}}{\varepsilon_2}} \right) ^{2}\nnr
&&\hspace{2mm}+\frac{{\it X_1} (x)}{ {\it \tilde{U}_1} (x) }  \left({\frac {{ 2\,\it r_h}\,\tilde{U}\,\rho\,{
\it d\tau}}{V \left( {{\it r_h}}^{2}+{x}^{2} \right) }}+{\frac {{a_1}^{2}\,{
\it \tilde{\gamma}_1} (x)\, {\it d\varphi_1}}{ \left( {a_1}^{2}-{x}^{2}
 \right) \varepsilon_1}}+{\frac {{a_2}^{2}\,{\it \tilde{\gamma}_2}
 (x)\, {\it d\varphi_2}}{ \left( {a_2}^{2}-{x}^{2} \right) \varepsilon_2
}} \right) ^{2}\,,
\eea
and its functions are
\bea
&&\tilde{U}=(x^2+a_1\,a_2)\,,\qquad  \Xi=1\,,\qquad V=4\,,\nnr
&&\tilde{\gamma}_1(x)=(a_1^2-x^2)\,(a_1+a_2)\,a_1\,,\qquad \varepsilon_1=a_1\,(a_1^2-a_2^2)\,,\nnr
&&\tilde{\gamma}_2(x)=(a_2^2-x^2\,)\,(a_1+a_2)\,a_2\,,\qquad  \varepsilon_2=-a_2\,(a_1^2-a_2^2)\,,\nnr
&&\tilde{U}_1(x)=-(x^2+a_1\,a_2)\,,\quad X_1(x)=x^{-2}\,(a_1^2-x^2)\,(a_2^2-x^2)\,.
\eea
As discussed in section \ref{KT-EVH}, this geometry has two second-rank Killing tensors. One of them, $K_{(0)}$, is the metric itself, and another is
\bea\label{K1}
K_{(1)}=&&{\frac {-4\,{x}^{2}}{ \left( a_1\,a_2+{x}^{2} \right) {\rho}^{2}}}\left(\partial_\tau\right)^2+{\frac {4\,{\rho}^{2}{x}^{2}}{a_1\,a_2+{x}^{2}}}\left(\partial_{\rho}\right)^2+{\frac { \left( {a_1}^{2}-{x}^{2} \right)  \left( {a_2}^{2}-{x}^{2}
 \right) a_1\,a_2}{{x}^{2} \left( a_1\,a_2+{x}^{2} \right) }}\left(\partial_{x}\right)^2\nnr
&&-{\frac {a_2 \left( {a_1}^{3}{a_2}^{2}+{a_1}^{3}{x}^{2}-2\,{a_1}^{2}{a_2}^{3}+4\,{
a_1}^{2}a_2{x}^{2}-a_1{a_2}^{4}-a_1{a_2}^{2}{x}^{2}-2\,a_2{x}^{4} \right) }{ \left( 
{a_1}^{2}-{x}^{2} \right)  \left( a_1+a_2 \right) ^{2} \left( a_1\,a_2+{x}^{2}
 \right) }}\left(\partial_{\varphi_1}\right)^2\nnr
&&+ {\frac {a_1 \left( {a_1}^{4}a_2+2\,{a_1}^{3}{a_2}^{2}-{a_1}^{2}{a_2}^{3}+{a_1}^{2}a_2{x}
^{2}-4\,a_1{a_2}^{2}{x}^{2}+2\,a_1{x}^{4}-{a_2}^{3}{x}^{2} \right) }{ \left( {a_2}^{2}-{x}^{2} \right)  \left( a_1+a_2 \right) ^{2} \left( a_1\,a_2+{x}^{2} \right) }}\left(\partial_{\varphi_2}\right)^2\nnr
&&+{\frac {2\,\sqrt {a_2}{x}^{2}}{\sqrt {a_1} \left( a_1\,a_2+{x}^{2} \right) \rho}}\,\left(\partial_\tau\right)\,\left(\partial_{\varphi_1}\right)+{\frac {2\,\sqrt {a_1}{x}^{2}}{\sqrt {a_2} \left( a_1\,a_2+{x}^{2} \right) \rho}}\,\left(\partial_\tau\right)\,\left(\partial_{\varphi_2}\right)\nnr
&&+{\frac {2\,{a_1}^{2}{a_2}^{2}-{a_1}^{2}{x}^{2}-{a_2}^{2}{x}^{2}}{ \left( a_1+a_2
 \right) ^{2} \left( a_1\,a_2+{x}^{2} \right) }}
\left(\partial_{\varphi_1}\right)\,\left(\partial_{\varphi_2}\right)\,.
\eea
The angular velocities are 
\bea
\Omega^1=\Omega^2=\frac{1}{(a_1+a_2)}\,.
\eea
Then, $b_0$ and the $b_i$'s in Eq. \eqref{b-comp} are equal to
\bea
&&b_0=-\frac{1}{2\,\Xi}\left(g^2\,r^2\,x^2+x^2-r^2\right)\,,\nnr
&&b_1=\frac{1}{2\,\varepsilon_1}\left[a_1^2\,r^2\,x^2+a_1^4\,(x^2-r^2)\right]\,,\nnr
&&b_2=\frac{1}{2\,\varepsilon_2}\left[a_2^2\,r^2\,x^2+a_2^4\,(x^2-r^2)\right]\,.
\eea
By applying the near-horizon transformation, 
\bea
r=r_h+\lambda\,r_h\, \rho\,,\quad dt=\beta\,\frac{d\tau}{\lambda}\,,\quad d\phi_i=d\varphi_i+\Omega^i\,dt\,,\quad \beta=\frac{(a_1+a_2)^2}{4\,r_h}\,,
\eea
to the principal potential $b$ in Eqs. \eqref{b} and \eqref{b-comp}, it is easy to see that 
\bea
\left(b_0+\Omega^1\,b_1+\Omega^2\,b_2\right)\big|_{r_h}=\frac{a_1^2\,a_2^2}{2\,(a_1+a_2)^2}\,,
\eea
which is constant. Therefore, choosing $C=-\frac{a_1^2\,a_2^2}{2\,(a_1+a_2)^2}$ will remove the divergent term of $b$ in the near-horizon limit. Then, we get  
\bea
\tilde{b}=\tilde{b}_0\,\rho\,d\tau+\tilde{b}_1\,d\varphi_1+\tilde{b}_2\,d\varphi_2\,,
\eea
where
\bea
&&\tilde{b}_0\equiv\beta\,r_h\,\frac{d}{dr}\left(b_0+\Omega^1\,b_1+\Omega^2\,b_2 \right)\big|_{r_h}=\frac{\sqrt{a_1\,a_2}\,(x^2+a_1\,a_2)}{4}\,.\nnr
&&\tilde{b}_1=-{\frac {{a_1}^{2} \left( {a_1}^{2}a_2-a_1{x}^{2}-{x}^{2}a_2 \right) }{2\,({a_1}^{2}
-{a_2}^{2})}}
\,,\qquad \tilde{b}_2={\frac {{a_2}^{2} \left( a_1{a_2}^{2}-a_1{x}^{2}-{x}^{2}a_2 \right) }{2\,({a_1}^{2}
-{a_2}^{2})}}\,.
\eea
The principal tensor, $h=d\, b$, reads
\bea
h=-\frac{\sqrt {a_1\,a_2}}{4}\, \left( a_1\,a_2+{x}^{2} \right) \,d\tau \wedge d\rho-\frac{\rho\,x\sqrt {a_1\,a_2}}{2}d\tau \wedge dx+\frac{x}{(a_1-a_2)}dx\wedge (a_1^2\, d\varphi_1-a_2^2\, d\varphi_2)\,.\nnr 
\eea
The Hodge dual of $h$ gives a Killing-Yano tensor of this form:
\bea
\star h= &&\hspace{-4mm}{\frac { \left( a_1+a_2 \right) x\,a_1\,a_2}{a_1-a_2}}\,d\tau \wedge d\rho \wedge d\varphi_1 - {\frac { \left( a_1\,a_2+{x}
^{2} \right)  \left( {a_1}^{2}-{x}^{2} \right) \sqrt {a_1\,a_2}}{4(a_1-a_2)}}\,d\tau \wedge d\rho \wedge d\varphi_2\nnr
&&\hspace{-4mm}+{\frac { \left( a_1+a_2 \right) \rho\, {a_1}\sqrt {a_1\,a_2}x}{2(a_1-a_2)}}d\tau \wedge dx \wedge d\varphi_1-{\frac { \left( a_1+a_2 \right) \rho \, a_2\,\sqrt {a_1\,a_2}x}{2(a_1-a_2)}}d\tau \wedge dx \wedge d\varphi_2\nnr
&&\hspace{-4mm}+ {\frac { \left( a_1\,a_2+{x}^{2} \right)  \left( {a_2}^{2}-{x}^{2} \right) 
\sqrt {a_1\,a_2}}{4(a_1-a_2)}} dx \wedge d\varphi_1 \wedge d\varphi_2\,.
\eea

The Killing tensor made of this Killing-Yano tensor, using Eq. \eqref{KT-KY}, is proportional to $K_{(1)}$ given by Eq. \eqref{K1}.

\subsection{Generic odd dimensions}\label{app-b1-2}
As discussed in section \ref{KT-extremal}, the principal potential, $b$, is divergent in the near-horizon limit. However, it is defined up to a shift of the form
\bea
b\to b+C_\mu\,dx^\mu\,,
\eea
with constant $C_\mu$'s. This shift does not affect $h=d\,b$. We use this freedom to make $b$ finite in the near-horizon limit.
The divergent term arises from $dt\to d\tau/\lambda$. In the following, we show that its coefficient, which is equal to $\left({b}_0+\sum_i {b}_i\,\Omega^i\right)\big|_{r_h}$, is a \textit{constant}. We start by substituting $b_0$ and the $b_i$'s from Eq. \eqref{b-comp}:
\bea
B\equiv \bigg({b}_0+\sum_{i=1}^n {b}_i\,\Omega^i\bigg)\bigg|_{r_h}
&=&-\frac{(r_h^2+a_0^2)}{2}\,\sum_{J=0}^{n}\left(\frac{\sum_{k=0}^{n-1}\,(-a_J^2)^{n-k}\,A^{(k+1)}}{(r_h^2+a_J^2)\,{\prod_{l=1}^{n}\,}'(a_i^2-a_l^2)}\right)\,.
\eea
The summation over $k$ can be easily done by changing $k=u-1$:
\bea
S_1\equiv\sum_{u=1}^{n}(-a_J^2)^{n+1-u}\,A^{(u)}&=&(-a_J^2)\,\left[\sum_{u=0}^{n}(-a_J^2)^{n-u}\,A^{(u)} -(-a_J^2)^n\,A^{(0)}\right]\nnr
&=&\prod_{\mu=0}^{n}\,(y_\mu^2-a_J^2)-(-a_J^2)^{n+1}\,,
\eea
where in the last line we used the definition $y_0^2\equiv 0$. The contribution of the last expression of $S_1$ to $B$ is a constant, $B_0$, which is desirable. Therefore, by applying the change
\bea
&&I=M-1\,, \qquad a_I=d_M\,,\nnr
&&\mu=\rho-1\,, \ \ \qquad y_\mu=z_\rho\,,
\eea
to the $B$, we have
\bea
B-B_0=\frac{(r_h^2+a_0^2)}{2}\,\sum_{M=1}^{n+1}\left[\frac{\prod_{\rho=1}^{n+1}(z_\rho^2-d_M^{\,2})}{(z_{n+1}^2-d_M^{\,2})}\right]\,\frac{1}{{\prod_{N=1}^{n+1}\,}' (d_M^{\,2}-d_N^{\,2})}\,.
\eea
Then, using the identity \eqref{expanded}, the expression in the bracket can be expanded in powers of $(-d_M^{\,2})$. Using the identity \eqref{n-2} for the summation over $M$, simplifies $B$ significantly and leads to
\bea
B-B_0=\frac{(r_h^2+a_0^2)}{2}\,,
\eea
which is obviously constant and can be absorbed by $C_0$. Therefore, the near-horizon expansion of $b_\tau$ starts from $\rho$, as explained in Eq. \eqref{NH-expansion}, and gives
\bea
\tilde{b}=\tilde{b}_0\,\rho\,d\tau+\sum_{i=1}^{n}\tilde{b}_i\, d \varphi_i\,,
\eea
where
\bea
\tilde{b}_0\equiv \beta\,r_h\,\sum_{k=0}^{n-1}\left[\frac{\Omega^i}{\varepsilon_i}(-a_i^2)^{n-k}-\frac{(-1)^n}{\Xi}(-g^2)^k\right]\tilde{A}^{(k)}_n\,.
\eea

\section{Useful identities} \label{app-b}
\bea\label{Ids}
&&\sum_{\a=1}^{N}\frac{\left(-x_\a^2\right)^{N-1-q}}{{\prod_{\beta=1}^{N}\,}'(x_\beta^2-x_\a^2)}=\delta_{q,0}\,,\label{n-2}\\ \nnr
&&\sum_{\a=1}^{N}\frac{A_\a^{(p)}}{x_\a^2\,{\prod_{\beta}\,}'(x_\beta^2-x_\a^2)}=\frac{A^{(p)}}{A^{(N)}}\,,\label{28}\\ \nnr
&&\frac{1}{\prod_{\a=1}^{N} (x_\a^2+\lambda)}= \sum_{\a=1}^{N} \frac{1}{{\prod_{\beta}}'(x_\beta^2-x_\a^2)}\,\frac{1}{(x_\a^2+\lambda)}\,,\label{27}\\ \nnr
&&\sum_{\a=1}^{N}\frac{A_\a^{(p)}\,(-x_\a^2)^{N-1-q}}{{\prod_{\beta}\,}'(x_\beta^2-x_\a^2)}=\delta_{q,p}\,,\qquad q=1,\ldots N-1\,, \label{29}\\ \nnr
&&\prod_{\beta=1}^N\, (x_\beta^2+\lambda)=\sum_{k=0}^{N}A^{(k)}\,\lambda^{N-k} \,, \qquad \frac{\prod_{\beta=1}^N \, (x_\beta^2+\lambda)}{(x_\alpha^2+\lambda)}=\sum_{k=0}^{N-1}\, A^{(k)}_{\alpha}\,\lambda^{N-1-k}\,.\label{expanded}
\eea

\end{document}